\pdfoutput=1
\documentclass[10pt,conference,letterpaper]{IEEEtran}

\usepackage[T1]{fontenc}
\usepackage[utf8]{inputenc} 
\usepackage{color}
\usepackage{soul}
\usepackage{ulem}
\usepackage{times,amsmath,epsfig}
\usepackage{multirow}
\usepackage{algpseudocode}

\makeatletter
\renewcommand{\ALG@beginalgorithmic}{\footnotesize}
\makeatother

\algtext*{EndWhile}
\algtext*{EndIf}
\algtext*{EndFor}

\title{Efficient XML Keyword Search based on DAG-Compression}
\author{%
{Stefan B\"ottcher{\small $~^{1}$}, Rita Hartel{\small $~^{2}$}, Jonathan
Rabe{\small $~^{3}$} }
\vspace{1.6mm}\\
\fontsize{10}{10}\selectfont\itshape
Institute for Computer Science, University of Paderborn\\
F\"urstenallee 11, 33102 Paderborn, Germany\\
\fontsize{9}{9}\selectfont\ttfamily\upshape
$~^{1}$stb@uni-paderborn.de\\
$~^{2}$rst@uni-paderborn.de\\
$~^{3}$jonny@mail.uni-paderborn.de%
}
\begin{document}
\maketitle
\begin{abstract} 
In contrast to XML query languages as e.g. XPath which require knowledge
on the query language as well as on the document structure, keyword search
is open to anybody. As the size of XML sources grows rapidly, the need for
efficient search indices on XML data that support keyword search increases. In
this paper, we present an approach of XML keyword search which is based on
the DAG of the XML data, where repeated substructures are considered only once,
and therefore, have to be searched only once. As our performance evaluation
shows, this DAG-based extension of the set intersection search
algorithm\cite{setintersection,setintersection2}, can lead
to search times that are on large
documents more than twice as fast as the search times of the XML-based
approach. Additionally, we utilize a smaller index, i.e., we consume less main
memory to compute the results.
\end{abstract}

\section{Introduction}
\subsection{Motivation}

The majority of the data within the internet is available nowadays in form
of tree-structured data (i.e. HTML, XML or 'XML dialects'). When searching for
certain information in huge document collections, the user typically (1) has no
knowledge of the structure of the document collection itself and
(2) is a non-expert user without any technical knowledge of XML or
XML query languages.
These requirements are met by XML keyword search where the user specifies the
searched information in form of a list of keywords (i.e., neither knowledge
of the document structure nor of any specific query language is required) and
document fragments are returned that contain each keyword of the specified keyword list.
For this purpose, the need for
efficient keyword search appraoches on XML data is high.

\subsection{Contributions}
Our paper presents \textit{IDCluster}, an approach to efficient keyword
search within XML data that is based on a DAG representation of the XML data,
where repeated substructures exist only once and therefore have to be searched
only once.
IDCluster combines the following features and advantages:
\begin{itemize}
\item Before building the index, IDCLuster removes redundant sub-trees and
splits the document into a list of so-called \textit{redundancy
components}, such that similar sub-trees have to be indexed and searched
only once.
\item For keyword search queries where parts of the results are contained
partially or completely within repeated sub-trees, the DAG-based keyword search
 outperforms the XML keyword search by a factor of more
than two on large documents, whereas it is comparably
fast for keyword search queries where all results occur in sub-trees that exist only once within the document.
\end{itemize} 
To the best of our knowledge, IDCluster is the first approach that shows these
advantages.
 
\subsection{Paper organization}

The rest of the paper is organized as follows. Section~\ref{sec:prel} introduces the underlying data model
and presents the ideas of the set intersection keyword
search algorithm \cite{setintersection}\cite{setintersection2} on which our
approach is based. Section~\ref{sec:ourapproach} presents our adaptation of the ideas of Zhou et al. to build the index
based on the document's DAG instead of the document's tree in order to avoid
repeated search within identical structures. Section \ref{sec:eval} contains
the performance evaluations, Section \ref{sec:relatedwork}
discusses the advantages of our approach to already existing approaches, and
finally, section \ref{sec:conclusions} concludes the paper with a short summary.

\section{Preliminaries}\label{sec:prel}

\subsection{Data model}

\begin{figure}[htb]
	\centering
	\includegraphics[width=\linewidth]{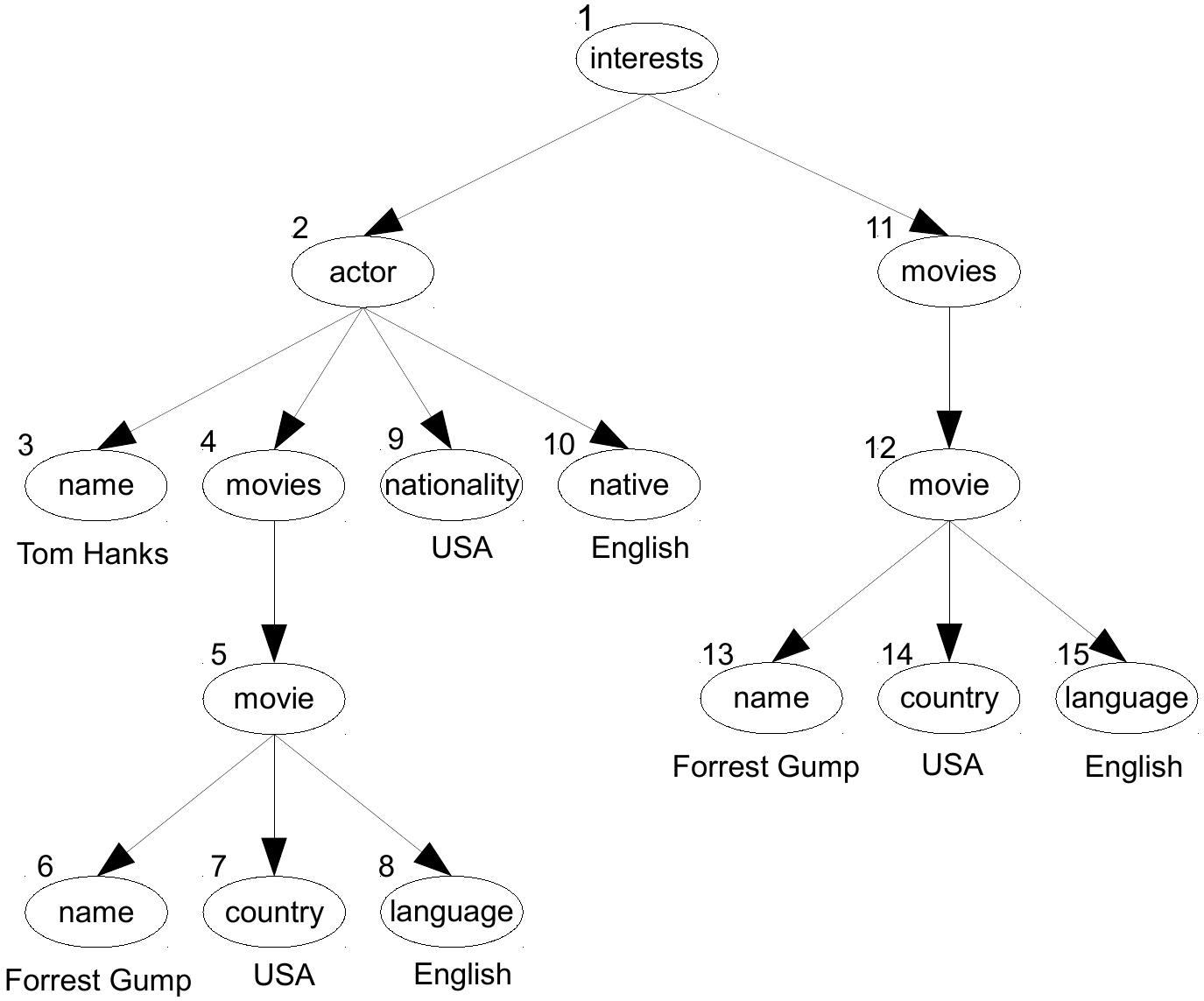}
	\caption{Example of an XML tree.}
	\label{fig:example}
\end{figure}

We model XML trees as conventional labeled ordered trees. Each node represents
an element or an attribute, while each edge represents a direct nesting
relationship between two nodes. We store a list of all keywords
$k_n$ contained in the element name or attribute name respectively, or
in any text value for
each node $n$. Thereby, we tokenize each text label into keywords at its
white-space characters. E.g., a node with label "name" and text value "Tom
Hanks" is split into the three keywords \textit{name}, \textit{Tom} and
\textit{Hanks}. The keywords in $k_n$ are called \textbf{directly contained}
keywords of $n$. Keywords that are directly contained in a descendant of $n$
are called \textbf{indirectly contained} keywords of $n$. Keywords that are
either directly or indirectly contained in $n$ are called \textbf{contained}
keywords of $n$. Furthermore, we assign each node its
pre-order traversal number assigned as ID. Fig.~\ref{fig:example} shows
a sample XML tree. The ID of each node is written top left next to it. The
element or attribute name is given inside the node's ellipse, while the text
values are given below.

\subsection{Query semantics}

In the last decade, different search goals for XML keyword search
queries have been proposed. These search goals form subsets of the common
ancestor (CA) nodes. For a query $Q$, the set of common ancestors $CA(Q)$ contains all nodes that
contain every keyword of $Q$. For the given query $Q_{ex}=\{USA,
English\}$ and the example tree of Fig.~\ref{fig:example}, the common ancestors are
$C(Q_{ex})=\{1,2,4,5,11,12\}$.
The two most widely adopted CA subsets are the smallest lowest common ancestor (SLCA) and
the exclusive lowest common ancestor (ELCA). All nodes in $CA(Q)$ that do not
have descendants in $CA(Q)$ are in $SLCA(Q)$, e.g. $SLCA(Q_{ex})=\{5,12\}$. A
node $n$ is in $ELCA(Q)$, when $n$ contains each
keyword of $Q$ outside of each subtree of a
descending CA node of $n$, e.g.,
$ELCA(Q_{ex})=\{2,5,12\}$. Node 2 is in $ELCA(Q_{ex})$, because when we would remove node 5
(which is CA) and its descendants, node 2 still contains both keywords:
\textit{USA} in node 9 and \textit{English} in node 10. Note that the given
definitions imply $SLCA(Q) \subseteq ELCA(Q) \subseteq CA(Q)$.

\subsection{IDLists}

This paper's algorithm is based on the set intersection keyword search
algorithm FwdSLCA and its modifications as proposed by Zhou et al.
\cite{setintersection,setintersection2}.

\begin{figure}[h]
	\centering
	\includegraphics[width=0.8\linewidth]{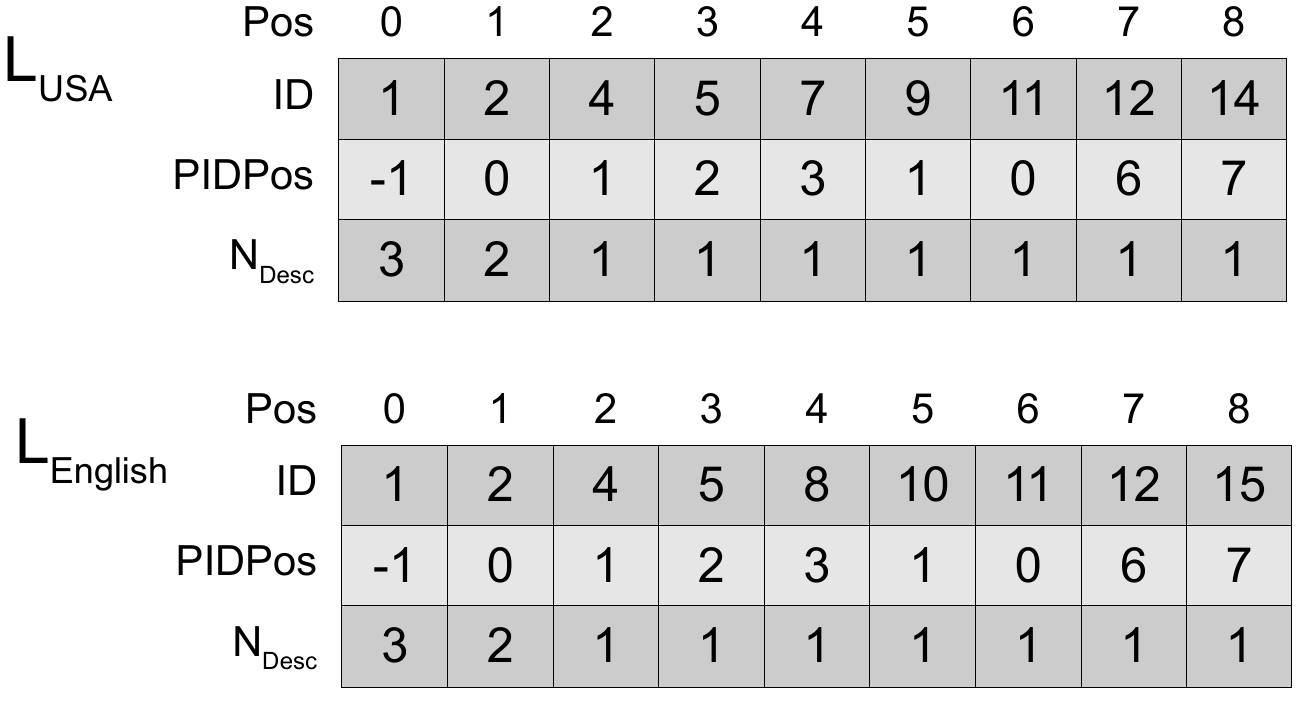}
	\caption{IDLists for the keywords \textit{USA} and \textit{English}.}
	\label{fig:idlists}
\end{figure}

Zhou et al. use IDLists as an index to efficiently perform keyword search. An
IDList is an inverted list of nodes that contain (directly or indirectly) a
certain keyword. For each node $n$ the IDList provides three values:
The \textit{ID} of $n$, the position of $n$'s parent inside the IDList named
$PIDPos$, and the number $N_{Desc}$ of nodes within $n$'s subtree
directly containing the keyword. IDLists are sorted by
ID.
Fig.~\ref{fig:idlists} shows the IDLists $L_{USA}$ and $L_{English}$ of
the keywords \textit{USA} and \textit{English} respectively. IDLists can be easily generated by
a single pass through the document.

\subsection{Search}\label{sec:otherapproach}

The general idea is to use set intersection on the IDLists of the query
keywords to find CA nodes. The found CA nodes are then checked against the SLCA or ELCA
semantics respectively to calculate the result set.

The basic SLCA search algorithm of \cite{setintersection,setintersection2},
\textit{FwdSLCA}, uses the method \textit{fwdGetCA} to efficiently calculate all CA nodes in
ascending order. This is done by maintaining pointers to the current position $C_i$ in
each IDList $L_i$, selecting the highest ID of all $C_i$, and binary searching
for this ID in the remaining IDLists. Due to the ascending order, a CA
node $n$ is SLCA if and only if the next found CA node is not a child of $n$.
This can be checked using the IDList's $PIDPos$.

For checking the ELCA semantics, the algorithm \textit{FwdELCA} uses an
additional stack for storing two arrays for each visited node $n$: One holds the
$N_{Desc}$ values of $n$, the other holds the accumulated $N_{Desc}$ values of
$n$'s CA children. As soon as every CA child of $n$ has been found, the
differences of those two arrays indicate whether $n$ is ELCA or not.

Beside \textit{FwdSLCA}, two more SLCA search algorithms based on set
intersection are proposed in \cite{setintersection,setintersection2}: The
algorithm \textit{BwdSLCA} finds CA nodes in reversed order, and as a result can
efficiently skip ancestors of CA nodes (note that ancestors of CA nodes by
definition never can be SLCA). The algorithm \textit{BwdSLCA+} additionally improves the binary search required for CA calculation by shrinking the search space. The shrinkage is based on the parent position information given by the $PIDPos$ value.

For \textit{FwdELCA}, one alternative is proposed: The algorithm
\textit{BwdELCA} finds CA nodes in reversed order and shrinks the binary search space like
\textit{BwdSLCA+}. The ancestor skipping introduced in \textit{BwdSLCA} cannot
be implemented for ELCA search since inner nodes can be an ELCA too.

%
%
%

\section{Our Approach}\label{sec:ourapproach}

The algorithm described in section~\ref{sec:otherapproach} is not
redundancy-sensitive, i.e., whenever there are repeated occurrences of
the same sub-tree, these occurrences are searched repeatedly.
However, the goal of our approach is to
follow the idea of DAG-based compression approaches and to exploit structural redundancies in
order to perform faster keyword search. This is done by splitting the original
XML tree into disjoint \textbf{redundancy components}. A
redundancy component is a subgraph of the original tree which occurs more than once within the tree. We then search
each redundancy component only once and combine the results to get the complete
result set, i.e., the same result as if we had performed the
search based on the tree and not based on the DAG.

\subsection{Index}

Our index is called IDCluster. It contains the IDLists for each redundancy
component and the redundancy component pointer map (RCPM) which is used for
combining the results of redundancy components. The IDCluster is generated in
two passes through the document.

\begin{figure}[htb]
	\centering
	\includegraphics[width=0.58\linewidth]{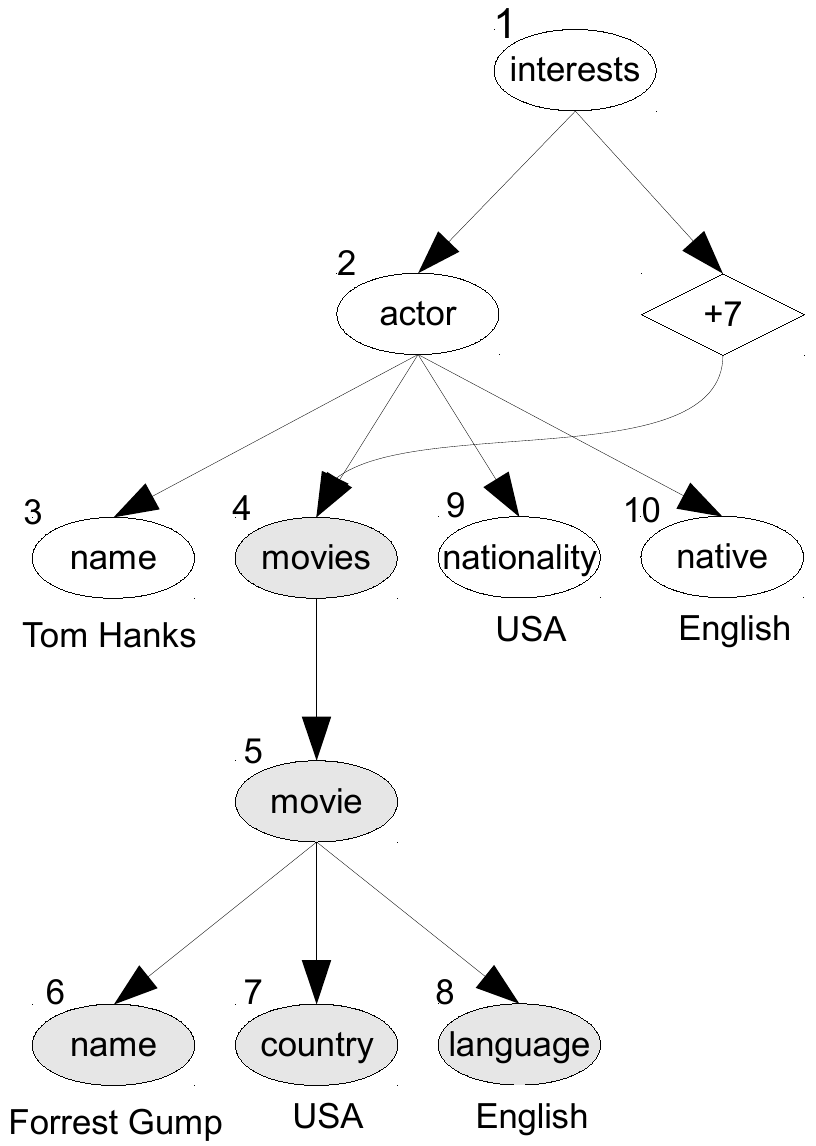}
	\caption{DAG compressed XML tree after the first traversal.}
	\label{fig:dag}
\end{figure}

The first pass is an extended DAG
compression where nodes are considered as being identical when (a) they directly contain the same keywords and (b) all
children are identical. When a node $n'$ is found which is identical to a node
$n$ found earlier, node $n'$ is deleted and the edge from its parent
$parent(n')$ to $n'$ is replaced by a new edge from $parent(n')$ to $n$. This new edge is called an offset edge and contains the difference between the IDs of
$n'$ and $n$ as an additional integer value. The information contained in offset
edges will later be used to recalculate the original node ID of $n'$.
Furthermore, for each node, we store the $OccurrenceCount$ which indicates the
number of identical occurrences of this node. Fig.~\ref{fig:dag} shows the XML
tree after the first traversal. The nodes with a white background have an $OccurrenceCount$ of 1, while nodes with a grey background have an $OccurenceCount$ of 2. E.g., node 5 is identical to node 12 in the original
tree. Hence node 12 is deleted and implicitly represented by node 5 and an
offset edge on its path with the offset +7. Node 5 has an $OccurrenceCount$ of
2, which indicates that node 5 now represents two nodes of the original tree.

\begin{figure}[htb]
	\centering
	\includegraphics[width=\linewidth]{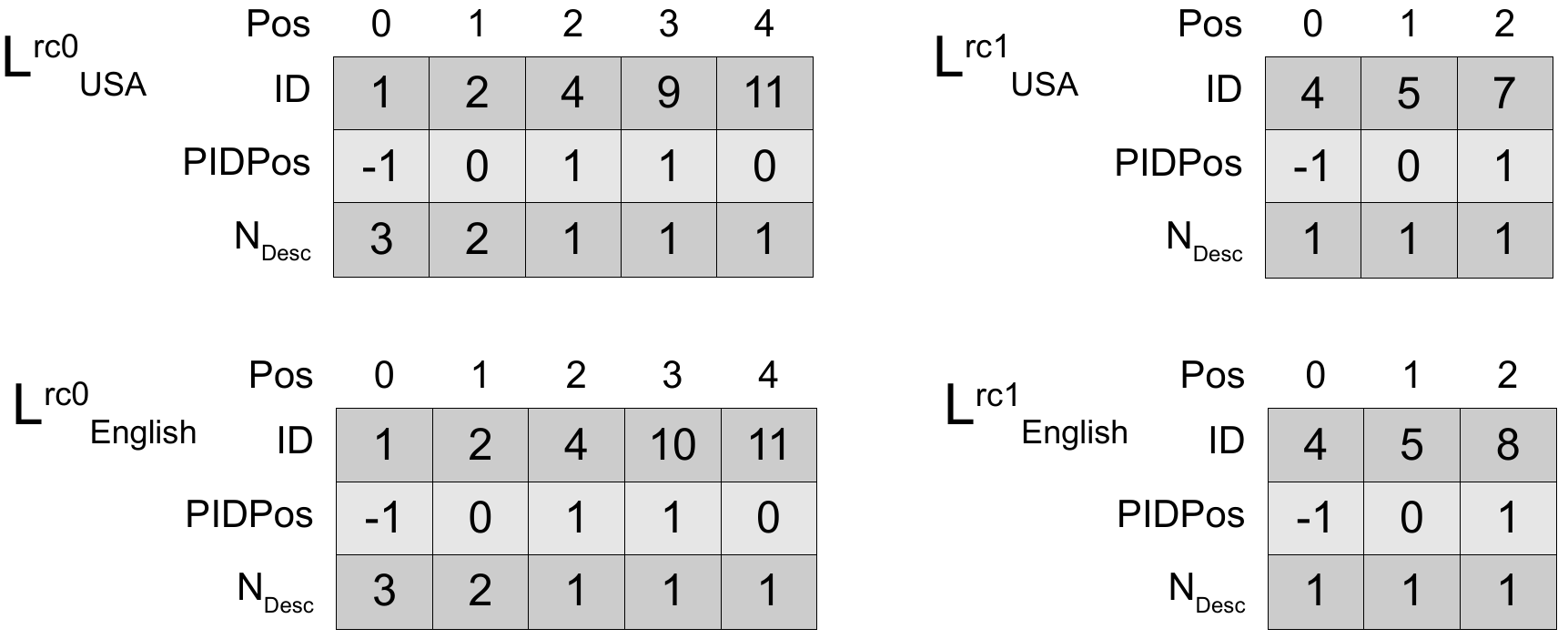}
	\caption{Created IDLists for the keywords \textit{USA} and \textit{English} as part of an IDCluster.}
	\label{fig:idcluster_idlists}
\end{figure}

In the second pass, the $OccurenceCount$ values are used for selecting
redundancy components: Each connected component consisting only of nodes with
the same $OccurenceCount$ is selected as a redundancy component. (The redundancy
component furthermore includes additional dummy nodes which represent nested
redundancy components and are introduced below.) Identifying redundancy
components and constructing their IDLists can be done easily in a single
document traversal utilizing the $OccurenceCount$. For each redundancy
component, a distinct set of IDLists is created. Each IDList entry for nodes
belonging to a redundancy component $rc$ is stored in the set of IDLists for
$rc$. The IDLists also include additional entries that represent nested
redundancy components. These additional entries, called dummy nodes, have the
same ID as the root node of the represented nested redundancy component
$rc_{nested}$ and are only added to IDLists of keywords contained in $rc_{nested}$.
Fig.~\ref{fig:idcluster_idlists} shows the IDLists created for the keywords
\textit{USA} and \textit{English}. Dummy nodes are the entries at the positions
2 and 4 in $L^{rc0}_{USA}$ and positions 2 and 4 in $L^{rc0}_{English}$.

\begin{figure}[htb]
	\centering
	\includegraphics[width=0.32\linewidth]{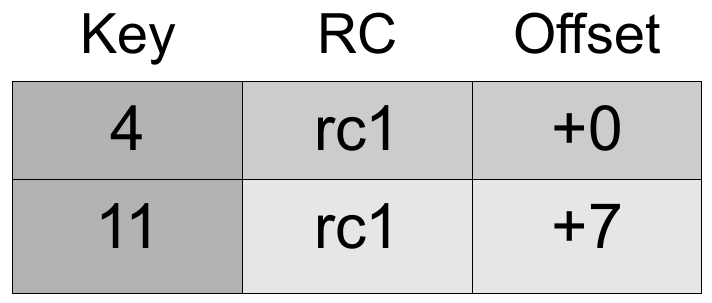}
	\caption{Created redundancy component pointer map (RCPM) as part of an IDCluster.}
	\label{fig:idcluster_rcpm}
\end{figure}

The dummy nodes are added to the redundancy component pointer map (RCPM). The
key of each entry in the RCPM is the ID of the respective dummy node. Each
entry contains the identifier of the redundancy component which the dummy node
is representing and an offset. The offset is given by the offset edge between $rc$
and $rc_{nested}$, or +0 if the edge between $rc$ and $rc_{nested}$ is not an
offset edge. Fig.~\ref{fig:idcluster_rcpm} shows the RCPM of the IDCluster. Note that only one RCPM is required, no matter how many keywords are indexed.

\subsection{Search}

\subsubsection{SLCA computation}

The main idea for SLCA computation is to utilize the base algorithms for
searching each redundancy component individually. Due to the additionally added
dummy nodes, any SLCA which is found in a redundancy component and is not a
dummy node, is also SLCA in the original tree. SLCA, which are dummy nodes,
indicate that more SLCA are inside the redundancy component, which the dummy
node is representing. This way, all redundancy components containing SLCA can be
searched dynamically starting at the root redundancy component. The offset
values stored in the RCPM are used for recalculating the original ID of SLCA
contained in nested redundancy components.

The basic search algorithm for IDCluster is an extension of FwdSLCA
\cite{setintersection2} and is depicted in Fig.~\ref{algo:dagfwdslca}. 
The algorithm starts searching in the redundancy component with the ID 0, which
by definition is the redundancy component containing the document's root
node. In lines \ref{line:startfwdslca}-\ref{line:endfwdslca} the procedure
DagFwdSLCA performs an SLCA search similar to FwdSLCA on one redundancy component.
In lines \ref{fwdslca:start}-\ref{fwdslca:end} the results from this redundancy
component are processed. To check whether an SLCA is a dummy node, the SLCA is
looked up in the RCPM in line \ref{fwdslca:checkrcpm}. If the SLCA is a dummy
node, more SLCA can be found in the nested redundancy component. Therefore, the
nested redundancy component is searched recursively in
line~\ref{fwdslca:recursion} if it was not searched before. The results from
the nested redundancy component are now modified by the offset and added to the
SLCA list in lines~\ref{fwdslca:replace}-\ref{fwdslca:addend}. Note that in
line~\ref{fwdslca:replace} the dummy node is deleted from the SLCA list by
replacing it with an actual SLCA contained in the nested redundancy component.
The only exception to this proceeding is when the SLCA list contains only the
root node of the current redundancy component. This is a special case since an
RCPM lookup would return dummy node information (the dummy node and the root
node of the nested redundancy component have the same ID) and cause an infinite loop.

\begin{figure}
\begin{algorithmic}[1]
\State \Call{DagFwdSLCA}{$IDCluster$,0}
\Statex 
\Procedure{DagFwdSLCA}{$IDCluster$,$rc_{cur}$}
\State $IDLists \longleftarrow$ $IDCluster$.getIDLists($rc_{cur}$)
 \While{$\neg$ EoL($IDLists$)}\label{line:startfwdslca} 
  \State $v \longleftarrow$ \Call{fwdGetCA}{$IDLists$}\label{line:fwdgetca}
  \If{$v \neq null$}
   \If{$u \neq null$ $\wedge$ $v$.parent $\neq u$}
    \State $SLCA[rc_{cur}]$.add($u$)
   \EndIf
   \State $u \longleftarrow v$
   \State \Call{advance}{$IDLists$}
  \EndIf
 \EndWhile
\If{$u \neq null$}
  $SLCA[rc_{cur}]$.add($u$)
\EndIf \label{line:endfwdslca}
\If{$SLCA[rc_{cur}][0] = IDLists[0].getID(0)$}\label{fwdslca:start}\label{fwdslca:root}
 \Return
\EndIf
\State $size \longleftarrow SLCA[rc_{cur}].$size
\For{$i\gets 0,size$}\label{fwdslca:startfor}
 \State $slca \longleftarrow SLCA[rc_{cur}][i]$
 \If{$slca \in IDCluster.RCPM$}\label{fwdslca:checkrcpm}
  \State $rc_{nes} \longleftarrow IDCluster$.getPointer($slca$)
  \State $os \longleftarrow IDCluster$.getOffset($slca$)
  \If{$\neg done[rc_{nes}]$}
   \State \Call{DagFwdSLCA}{$IDCluster$,$rc_{nes}$}\label{fwdslca:recursion}
  \EndIf
  \State $SLCA[rc_{cur}][i]$ $\longleftarrow$ $SLCA[rc_{nes}][0] + os$\label{fwdslca:replace}
  \ForAll{$slca_{nes} \in  SLCA[rc_{nes}] \setminus SLCA[rc_{nes}][0]$}\label{fwdslca:addstart}
   \State $SLCA[rc_{cur}]$.add($slca_{nes} + os$)
  \EndFor\label{fwdslca:addend}
 \EndIf
\EndFor \label{fwdslca:end}\label{fwdslca:endfor}
\State $done[rc_{cur}] \longleftarrow $true
\EndProcedure
\Statex 
\Function{EoL}{$IDLists$}
 \ForAll{$IDList \in IDLists$}
  \If{$IDList.C_i \geq IDList.length$}
   \State \Return true
  \EndIf
 \EndFor
 \State \Return false
\EndFunction
\Statex
\Procedure{advance}{$IDLists$}
 \ForAll{$IDList \in IDLists$}
  \State $IDList.C_i \longleftarrow IDList.C_i$+1
 \EndFor
\EndProcedure
\end{algorithmic}
\caption{Alternative SLCA search algorithm for redundancy components}\label{algo:dagfwdslca}
\end{figure}

If we consider our example, a search for the keywords \textit{USA}
and \textit{English} starts with the call of the main algorithm for the root
redundancy component $rc_0$. In lines \ref{fwdslca:start}-\ref{fwdslca:end},
node 4 and node 11 are calculated as the SLCA results for this redundancy
component. In the first iteration of the loop in lines
\ref{fwdslca:startfor}-\ref{fwdslca:endfor}, node 4 is identified as a dummy
node (line~\ref{fwdslca:checkrcpm}). As there are no results yet for the nested
redundancy component, $rc_1$ is searched recursively now. In $rc_1$, the only
SLCA result is node 5. Since node 5 is not found in the RCPM
(line~\ref{fwdslca:checkrcpm}), the SLCA result list for $rc_1$ remains
unchanged and the recursive call terminates. Back in the parenting recursion
for $rc_0$, inside the SLCA result list, the dummy node 4 is replaced with the
first result of the nested redundancy component, increased by the offset, 5+0=5
in line~\ref{fwdslca:replace}. Since no more SLCA results exist in $rc_1$, the
loop in lines~\ref{fwdslca:addstart}-\ref{fwdslca:addend} is skipped. In the
next iteration of the outer loop, node 11 is identified as a dummy node in
line~\ref{fwdslca:checkrcpm}. The nested redundancy component is once again
$rc_1$, for which at this point results already exist. Therefore, $rc_1$ is not
searched again and the dummy node is replaced by the SLCA result from $rc_1$
increased by the offset given by the RCPM, 5+7=12. At this point, the outer loop
terminates, and with it, the algorithm terminates, while the SLCA result list for $rc_0$ contains node 5 and node 12 as the final result.

One advantage of this approach is that the algorithm given by Zhou et al. is
integrated as unmodified module. This means our approach will benefit from any improvements made to the base algorithm, like the parent skipping introduced in BwdSLCA or the improved binary search introduced in BwdSLCA+.

\subsubsection{ELCA computation}

ELCA search can be implemented in a similar manner, which is shown in Figure~\ref{algo:dagfwdelca}.
Lines~\ref{fwdelca:same_start}-\ref{fwdelca:same_end} are similar to the
algorithm given by Zhou et al. \cite{setintersection2}. The following lines are
adopted from the DagFwdSLCA algorithm. Note that a redundancy component can
contain multiple ELCA, even when the root is ELCA. So, if
the first ELCA is the root, the first ELCA is just skipped in
line~\ref{fwdelca:skipfirst} instead of aborting the whole function call as in DagFwdSLCA.

\begin{figure}
\begin{algorithmic}[1]
\State \Call{DagFwdELCA}{$IDCluster$,0}
\Statex  
\Procedure{DagFwdELCA}{$IDCluster$,$rc_{cur}$}
\State $IDLists \longleftarrow$ $IDCluster$.getIDLists($rc_{cur}$)\label{fwdelca:same_start}
 \While{$\neg$ EoL($IDLists$)} 
  \State $v \longleftarrow$ \Call{fwdGetCA}{$IDLists$}
  \If{$v \neq null$}
   \While{$\neg S$.empty $\wedge$ $ S.$top $\neq v.$parent}
    \State \Call{ProcessStackEntry}{}
   \EndWhile  
   \State $S$.push($v$)
  \Else
   $ $ break
  \EndIf
  \State \Call{advance}{$IDLists$}
 \EndWhile
 \While{$\neg S$.empty}
  \State \Call{ProcessStackEntry}{}
 \EndWhile  \label{fwdelca:same_end}
\State $start \longleftarrow 0$
\If{$ELCA[rc_{cur}][0] = IDLists[0].getID(0)$}\label{fwdelca:start}\label{fwdelca:root}
 \State $start \longleftarrow 1$\label{fwdelca:skipfirst}  
\EndIf
\State $size \longleftarrow ELCA[rc_{cur}].$size
\For{$i\gets start,size$}
 \State $elca \longleftarrow ELCA[rc_{cur}][i]$
   \If{$elca \in IDCluster.RCPM$}
    \State $rc_{nes} \longleftarrow IDCluster$.getPointer($elca$)
    \State $os \longleftarrow IDCluster$.getOffset($elca$)
    \If{$\neg done[rc_{nes}]$}
     \State \Call{DAGFwdELCA}{$IDCluster$,$rc_{nes}$}
    \EndIf
    \State $ELCA[rc_{cur}][i]$ $\longleftarrow$ $ELCA[rc_{nes}][0] + os$\label{fwdelca:replace}
    \ForAll{$elca_{nes} \in  ELCA[rc_{nes}] \setminus ELCA[rc_{nes}][0]$}\label{fwdelca:addstart}
     \State $ELCA[rc_{cur}]$.add($elca_{nes} + os$)
    \EndFor\label{fwdelca:addend}    
   \EndIf
\EndFor
\State $done[rc_{cur}] \longleftarrow $true
\EndProcedure
\end{algorithmic}
\caption{Alternative ELCA search algorithm for redundancy components.}\label{algo:dagfwdelca}
\end{figure}

In a search for \textit{USA} and \textit{English}, the ELCA nodes found in
$rc_0$ are 2, 4 and 11. Node 2 is not in the RCPM and therefore a final ELCA.
Node 4 is a dummy node and forces a search in $rc_1$. The ELCA list in $rc_1$
contains node 5 only. Therefore, node 4 in the ELCA list of $rc_0$ is replaced
with 5 plus the offset 0. Node 11 is also a dummy node pointing to $rc_1$.
Since $rc_1$ was already searched, node 11 in the ELCA list of $rc_0$ is
replaced with 5 plus the offset 7. Therefore, the final ELCA results are 2, 5,
and 12.

\section{Evaluation}\label{sec:eval}

To test the performance of this paper's algorithms, comprehensive
experiments were run. The experiments focus on the comparison
between the base algorithms and their respective DAG variants introduced in this
paper. An evaluation of the base algorithms showing their
superior performance in comparison to other classes of keyword search
algorithms can be found in the original papers \cite{setintersection}\cite{setintersection2}.

\subsection{Setup}

All experiments were run on a Xeon E5-2670 with 256GB memory and Linux OS. The
algorithms were implemented in Java 1.6.0\_24 and executed using the OpenJDK
64-Bit Server VM. The time results are the averages of 1,000 runs
with warm cache.

The XML version of the music database
discogs.com\footnote{http://www.discogs.com/data/} is used as testdata. It
contains 4.2 million records of music releases of a size of 12.6GB.
To evaluate the effects of different database
sizes, smaller file sizes are created by successivly removing the second
half of the set of all remaining records. Thereby, additional databases
having the sizes of 0.8GB/1.6GB/3.3GB/6.5GB are created.

For the evaluation, 3 categories of querys are proposed:

\begin{itemize}
\item Category 1: Queries consisting of nodes that will not be compressed in a
DAG. DAG-based algorithms cannot exploit these
kinds of queries.
Since the DAG-based algorithms still need time for verifying the
absence of RC-Pointers (nodes with entries in the RCPM), they should have worse performance than the base algorithms.
\item Category 2: Queries consisting of nodes that will be compressed in a
DAG, but having common ancestors (CA) which still cannot
be compressed.
This means that all results will be in the first redundancy component.
DAG-based algorithms can exploit the fact that the IDLists for
the first redundancy can be shorter than the IDLists for the base algorithms. On
the other hand, the absence of all RC-Pointers still has to be
verified.
These advantages and disadvantages might cancel each other depending
on the situation.
\item Category 3: Queries with results that can be compressed in a
DAG.
This means that there has to be at least a second redundancy component
containing all keywords. DAG-based algorithms should have a
better performance than the base algorithms for queries from this category.
\end{itemize}

Queries of different lengths for all categories are selected randomly using the
200 most frequent keywords. The selected queries are shown in
Table~\ref{table:queries}. Table~\ref{table:queryproperties} shows
the properties of these queries. The columns \textbf{CA}, \textbf{ELCA} and
\textbf{SLCA} show the total number of CA-, ELCA- or SLCA-nodes respectively.
The columns \textbf{S$_{ca}$}, \textbf{S$_{elca}$} and \textbf{S$_{slca}$} show
the savings by DAG compression, e.g. a total number of 100 CA and
80\% CA savings imply that 20 CA are left in the XML tree after
DAG compression.
The properties of the used keywords can be found in
Table~\ref{table:keywords}.
The column \textbf{nodes} shows the number of nodes directly containing the respective keyword; the column \textbf{path} shows the number of nodes directly or indirectly containing the keyword. The columns S$_{nodes}$ and S$_{path}$ accordingly show the compression savings.

\begin{table}[htbp]
\begin{center}
\begin{tabular}{|c|c|c|l|}
\hline
\textbf{query}  & \textbf{category} & \textbf{length} & \multicolumn{1}{c|}{\textbf{keywords}} \\ \hline \hline
Q1 & \multirow{3}{*}{1} & 2 & image uri \\ \cline{1-1} \cline{3-4}
Q2 &  & 3 & image uri release \\ \cline{1-1} \cline{3-4}
Q3 &  & 4 & image uri release identifiers \\ \hline \hline
Q4 & \multirow{3}{*}{2} & 2 & vinyl electronic \\ \cline{1-1} \cline{3-4}
Q5 &  & 3 & vinyl electronic 12" \\ \cline{1-1} \cline{3-4}
Q6 &  & 4 & vinyl electronic 12" uk \\ \hline \hline
Q7 & \multirow{3}{*}{3} & 2 & description rpm \\ \cline{1-1} \cline{3-4}
Q8 &  & 3 & description rpm 45 \\ \cline{1-1} \cline{3-4}
Q9 &  & 4 & description rpm 45 7" \\ \hline
\end{tabular}
\end{center}
\caption{Queries for evaluation.}\label{table:queries}
\end{table}

\begin{table}[htbp]
\begin{center}
\begin{tabular}{|c|c|c|c|c|c|c|}
\hline
\textbf{query} & \textbf{CA } & \textbf{S$_{ca}$} & \textbf{ELCA } & \textbf{S$_{elca}$} & \textbf{SLCA } & \textbf{S$_{slca}$} \\ \hline \hline
Q1 & 14818739 & 0\% & 7697608 & 0\% & 7697603 & 0\% \\ \hline
Q2 & 3560569 & 0\% & 3560568 & 0\% & 3560567 & 0\% \\ \hline
Q3 & 1299279 & 0\% & 1299278 & 0\% & 1299277 & 0\% \\ \hline \hline
Q4 & 824063 & 0\% & 823952 & 0\% & 823840 & 0\% \\ \hline
Q5 & 616919 & 0\% & 616907 & 0\% & 616905 & 0\% \\ \hline
Q6 & 207865 & 0\% & 207864 & 0\% & 207863 & 0\% \\ \hline \hline
Q7 & 3438507 & 78\% & 709713 & 98\% & 703949 & 99\% \\ \hline
Q8 & 2389299 & 78\% & 486093 & 98\% & 481211 & 99\% \\ \hline
Q9 & 1307891 & 73\% & 328419 & 99\% & 328386 & 99\% \\ \hline
\end{tabular}\end{center}
\caption{Properties of queries.}\label{table:queryproperties}
\end{table}

\begin{table}[htbp]
\begin{center}
\begin{tabular}{|c|c|c|c|c|}
\hline
\textbf{keyword} & \textbf{path} & \textbf{S$_{path}$} & \textbf{nodes} & \textbf{S$_{nodes}$} \\ \hline \hline
image & 14862218 & 0\% & 7716896 & 0\% \\ \hline
uri & 22521466 & 0\% & 7699158 & 0\% \\ \hline
release & 4788936 & 9\% & 4447144 & 3\% \\ \hline
identifiers & 2845185 & 6\% & 1422641 & 12\% \\ \hline
vinyl & 9088129 & 72\% & 2343810 & 97\% \\ \hline
electronic & 5305014 & 66\% & 1773128 & 99\% \\ \hline
12" & 3976817 & 77\% & 812587 & 97\% \\ \hline
uk & 1753528 & 50\% & 872902 & 95\% \\ \hline
description & 36872585 & 76\% & 12774491 & 92\% \\ \hline
rpm & 3460698 & 78\% & 715987 & 98\% \\ \hline
45 & 2524740 & 76\% & 531299 & 95\% \\ \hline
7" & 3371487 & 77\% & 689983 & 97\% \\ \hline
\end{tabular}
\end{center}
\caption{Properties of keywords.}\label{table:keywords}
\end{table}

\subsection{Experiment I: Category}

\begin{figure}[htb]
	\centering
	\includegraphics[width=0.5\linewidth]{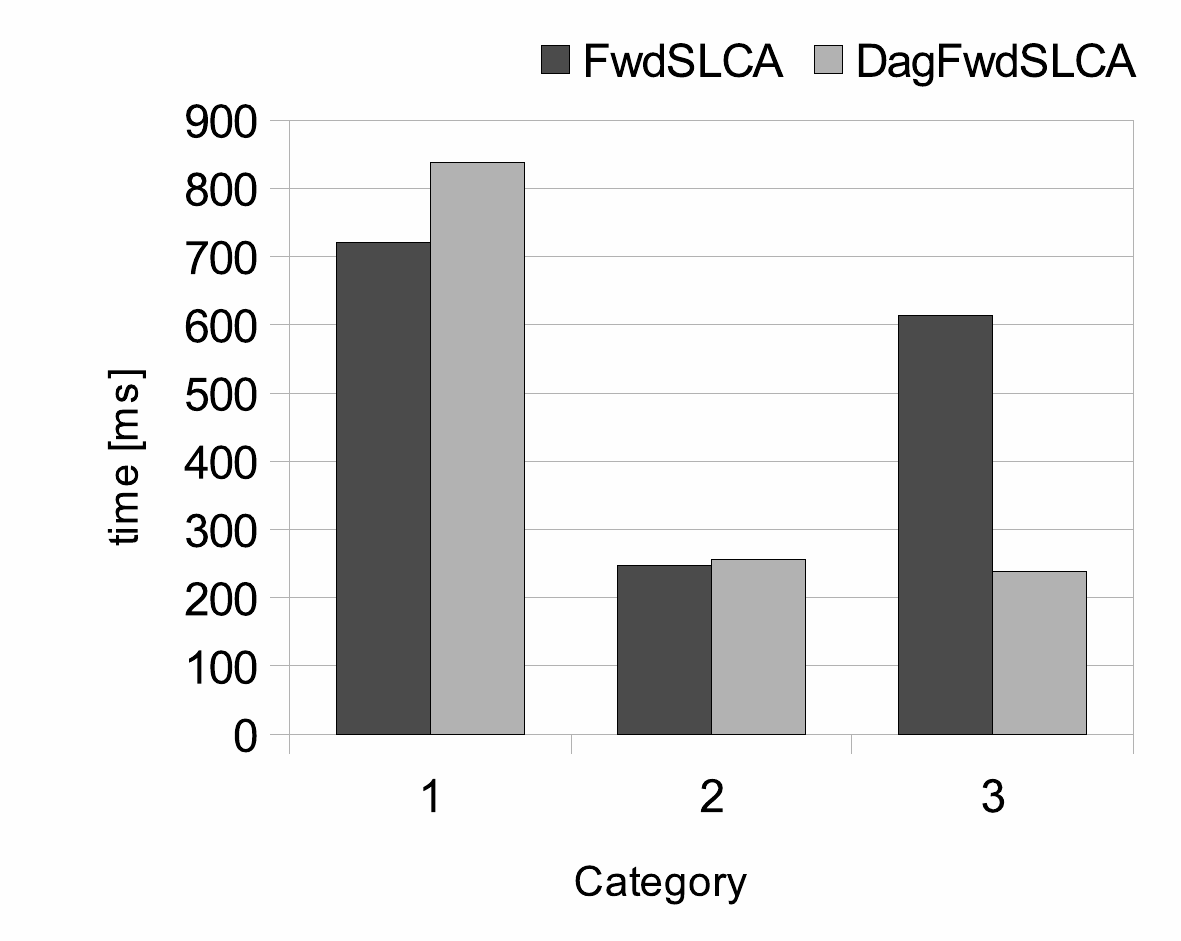}
	\caption{Comparison with different categories. (database size:
	12.6GB; query length: 3)}
	\label{eval:class}
\end{figure}

In the first experiment, the performance of the base algorithm FwdSLCA
is compared to the DAG-based algorithm DagFwdSLCA. Database size
and query length are fixed, while the category is altered. The results are shown
in Fig.~\ref{eval:class}.

The results confirm that in Category 1, the performance of the
DAG-based algorithm is a bit worse than the performance of the
base algorithm. This is as expected, since there are no redundancies which can
be exploited by a DAG-based algorithm, but
the DAG-based algorithm has a certain overhead for verifying that
no further redundancy components needs to be searched, resulting in a worse
performance. In Category 2, the performance of both algorithms is very
similar with the base algorithm being slightly faster. The better relative
performance of the DAG-based algorithm can be traced back to the
IDLists being shorter than the IDLists used in the base algorithm.
Finally, in Category 3, the DAG-based algorithm is more than
twice as fast as the base algorithm.

\subsection{Experiment II: Query length}

\begin{figure}[htb]
	\centering
	\includegraphics[width=\linewidth]{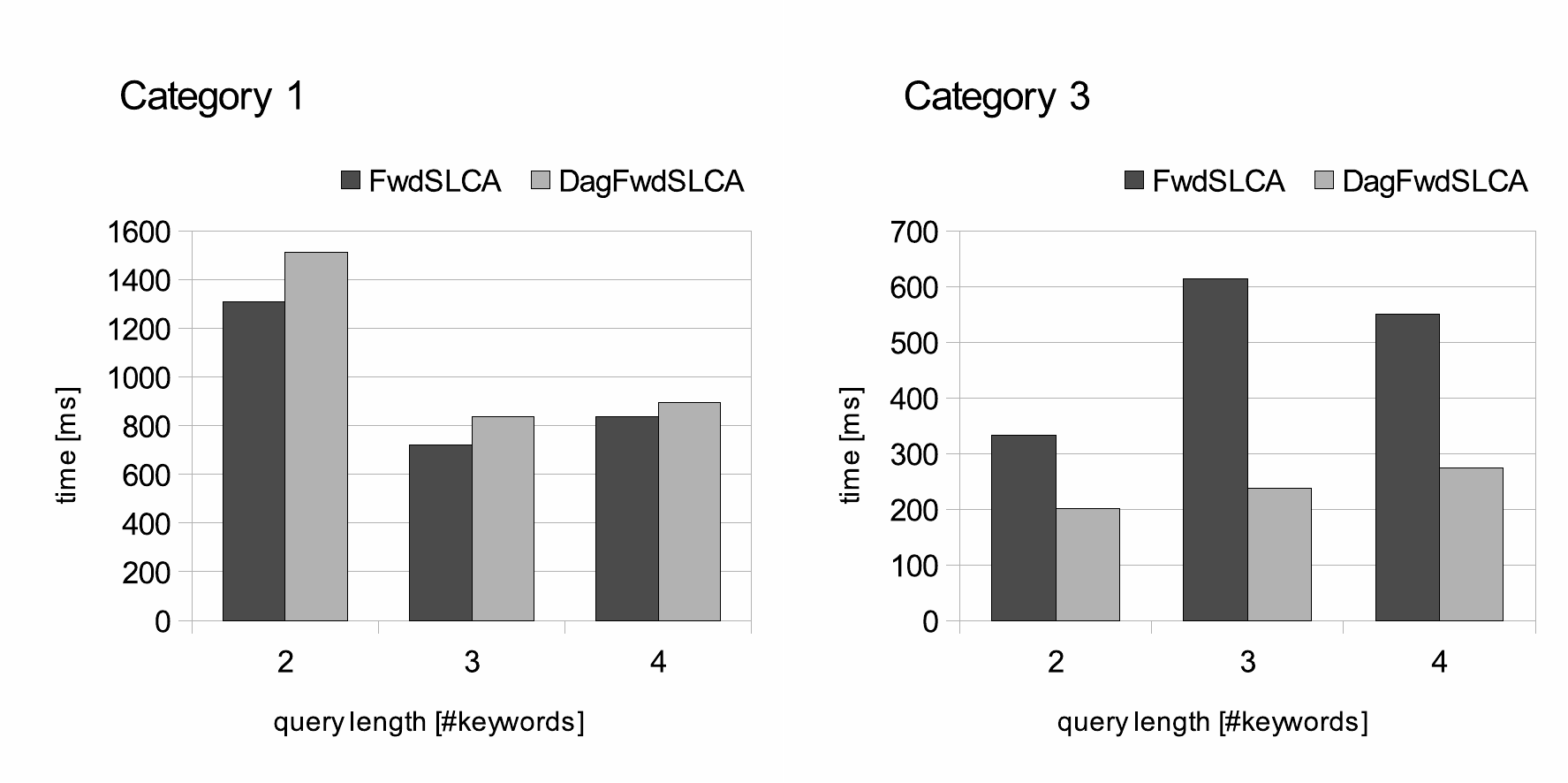}
	\caption{Comparison with different query lengths. (database size:
	12.6GB)}
	\label{eval:length}
\end{figure}

In this experiment, the length of the queries is modified.
Fig.~\ref{eval:length} shows the results for categories 1 and 3
with a fixed database size.

The general tendency is the same: The base algorithm is a bit better for
Category 1 queries, while the DAG-based algorithms are better for
Category 3 queries. The Category 1 results suggest that the gap between both
algorithms get smaller the more keywords are used. This is plausible, since the
overhead for the DAG-based algorithms depends on the amount of
results.
Adding more keywords to a query can reduce the amount of results, but never
increase it (see Table~\ref{table:queryproperties}).

\subsection{Experiment III: Database size}

\begin{figure}[htb]
	\centering
	\includegraphics[width=\linewidth]{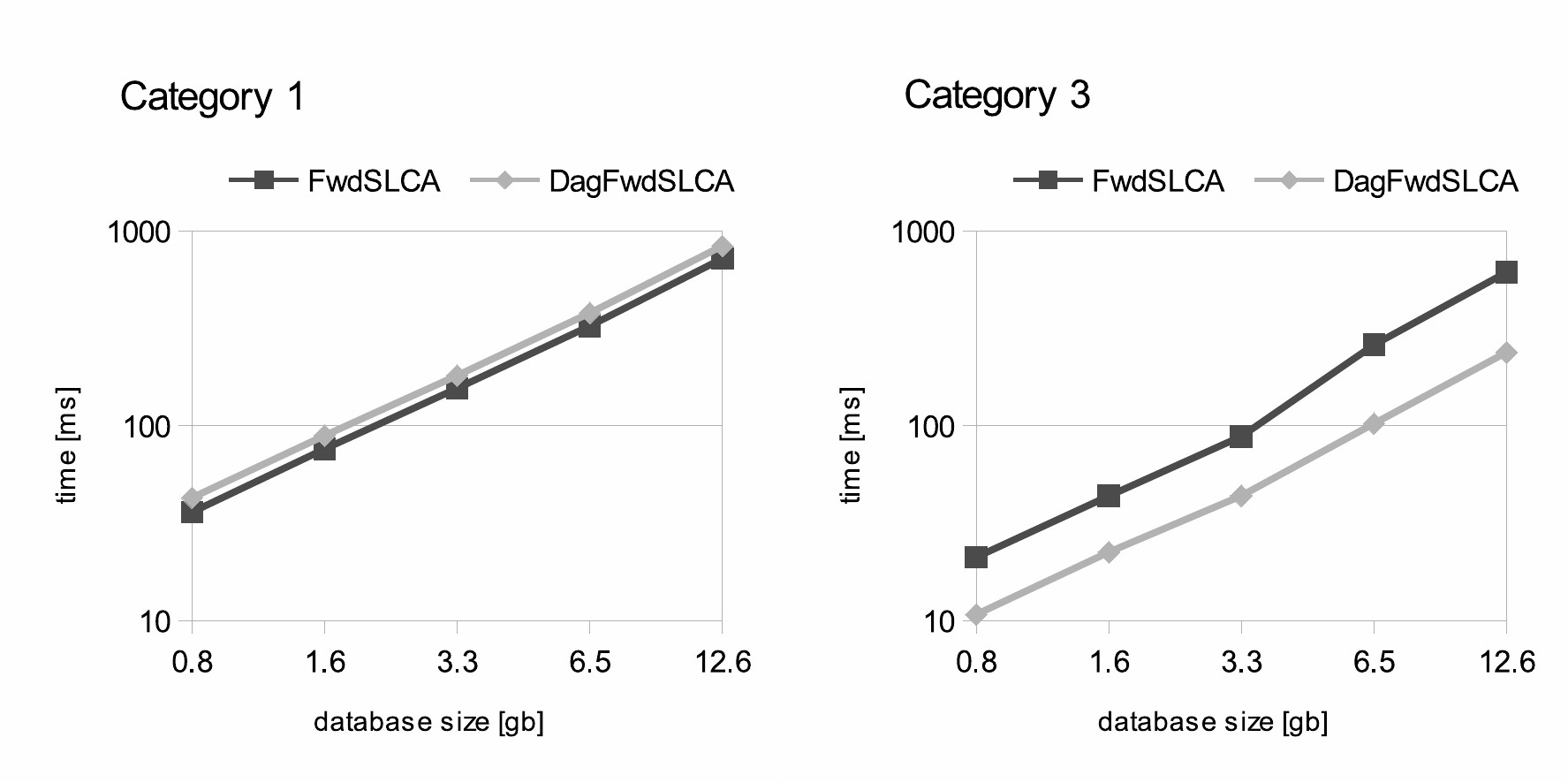}
	\caption{Comparison with different database sizes. (query length:
	3)}
	\label{eval:size}
\end{figure}

The third experiment examines the effects of database size and is shown in
Fig.~\ref{eval:size} for Category 1 and Category 3 with a fixed query
length.

The exponential growth of the database size leads to an exponential growth
in the search time for both algorithms. Minor changes in the proportions between
both algorithms can be traced back to minor changes in the keyword frequencies
and compression savings. Therefore, the database size seems not to have
a direct impact on the performance ratio between base algorithm and
DAG-based algorithm.

\subsection{Experiment IV: Algorithm}

In the last experiment, algorithms FwdSLCA, BwdSLCA+, FwdELCA and BwdELCA as
proposed by Zhou et al. are compared to their DAG-based variants.
Database size and query length are fixed.

\begin{figure}[htb]
	\centering
	\includegraphics[width=\linewidth]{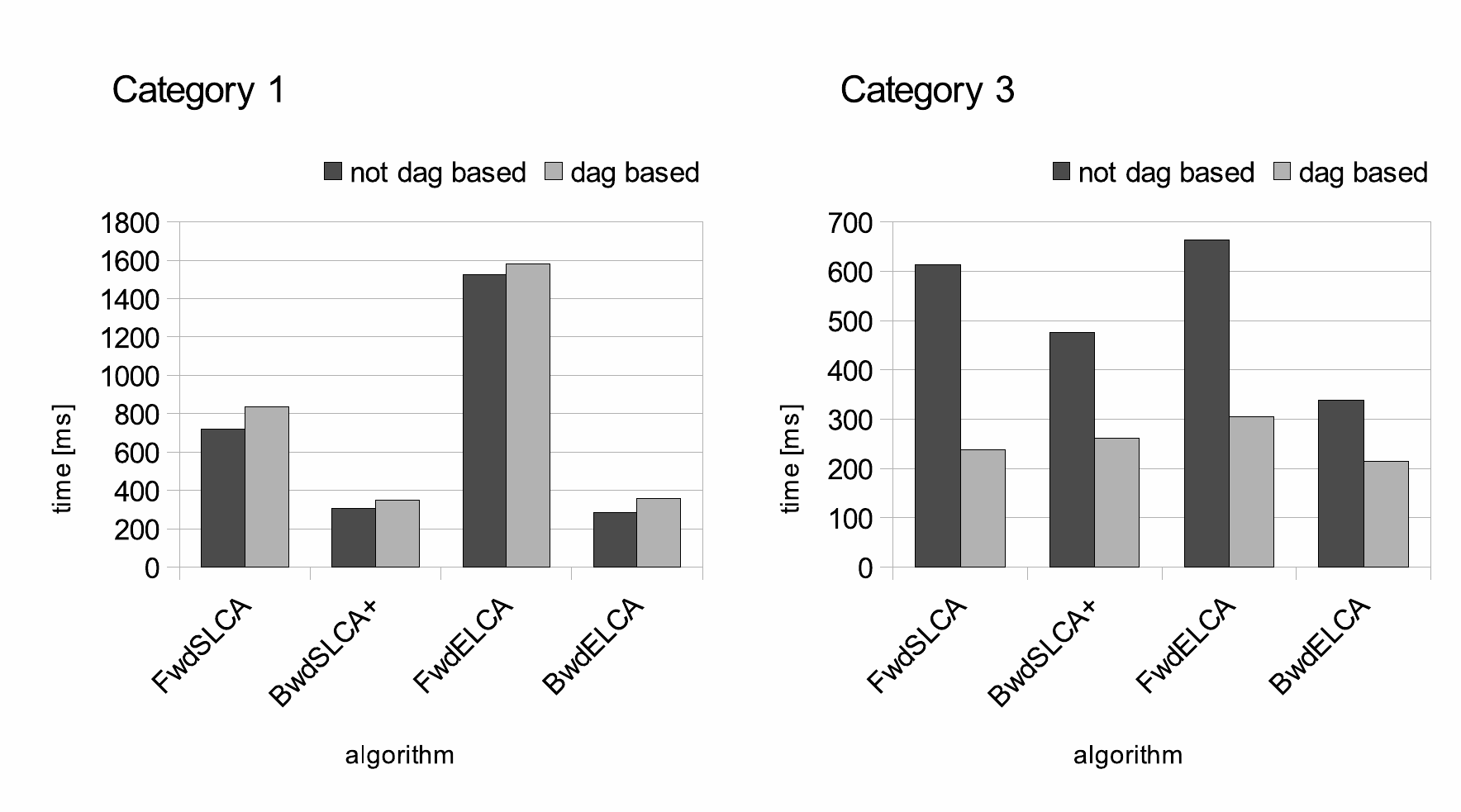}
	\caption{Comparison with different algorithms. (database size:
	12.6GB, query length: 3)}
	\label{eval:algorithms}
\end{figure}

For Category 1, the DAG-based algorithms always have a small
overhead independent of the algorithm. In Category 3, the
relative difference is smaller for backward algorithms, but still significant.
Backward search performs generally better than forward search of the same type.
The only exceptions are the DAG-based variants of
FwdSLCA and of BwdSLCA+ in Category 3. The backward algorithm is actually
slower.
This result suggests that a major part of the speedup in BwdSLCA+ is generated
by parent skipping. Due to the DAG compression many of the cases
in which parent skipping provides benefits are already optimized.

\subsection{Index Size}

The size of the IDCluster differs from the size of IDLists in two aspects.
On the one hand, additional space is required for storing the RCPM. On the other hand, less space is required for storing the IDLists due to DAG compression.

The RCPM can be stored in different ways which affect the required memory space and the time performance. In this evaluation, the RCPM is stored as an array containing the redundancy component identifier and the offset. The node ID is implicitly represented by the position in the array. The size of the array has to be big enough to contain all node IDs. This way of storing the RCPM is optimized for time performance.

The DAG compression strongly depends on the XML database used. The additionally created dummy nodes also have to be considered.

In a typical use case, both effects, additional memory for RCPM and
reduced memory due to DAG compression, are likely to cancel each other.

In the Discogs database, the total amount of nodes in
the IDLists index is 3.9 billion. Considering the 2 (3) integers per node
required for performing an SLCA (ELCA) search and an integer size of 4
byte, the total IDLists index sums up to 28.7GB (43.0GB). The amount of nodes in
an IDCluster is only 3.0 billion for the same database. Storing these nodes sums
up to 22.5GB (33.7GB) for SLCA (ELCA) search. However storing the RCPM in an array
as described above (with redundancy component identifier and offset both as 4
byte integer) for all 656 million distinct nodes requires an additional 4.9GB.
So, the total memory required for the IDCluster is 27.3GB (38.6GB) for SLCA (ELCA) search.

\section{Related Work}\label{sec:relatedwork}

There exist several approaches that address the problem of keyword search in
XML. These approaches can be roughly divided into two categories: approaches
that enhance the quality of the search results by considering the semantics of
the queries on the one hand, and approaches that enhance the performance
of the computation of the set of query results on the other hand.

Within the first category, XSEarch \cite{xsearch} presents a query
semantics that returns only those XML fragments, the result nodes of which
are \textit{meaningfully related}, i.e., intuitively belong to the same entity.
In order to check this, they examine whether a pair of result
nodes has two different ancestor nodes that have the same label (e.g., two nodes
with label ``author'', s.th. the first keyword belongs to author1
and the second one to author2).

\cite{xrank} not only focusses on an efficient, stack-based algorithm for
keyword search based on inverted element lists of the node's DeweyIDs, but also
aims to rank the search results in such a way, that the user
gets the (probably) most interesting results prior to the other results. SUITS \cite{demidovasuits} is a
heuristics-based approach, and the approach presented in
\cite{petkova2009refining} uses probabilistic scoring to rank the query results.
In order to enhance the usability, \cite{li2010suggestion} and
\cite{koutrika2009data} propose an approach on how to group the query results by
category.

Within the second category (efficient result computation) most approaches are
based on finding a set of SLCA (or ELCA) nodes for all
matches of a given keyword list.

Early approaches were computing the LCA for a set of given keywords on the fly.
\cite{schmidt2001querying} proposes the meet-operator that computes the
LCA for a pair of nodes that match two query strings without requiring additional
knowledge on the document structure from the user.

In contrast, recent approaches try to enhance the query performance by using a
pre-computed index.

\cite{florescu2000integrating} proposes an extension of the XML query
language XML-QL by keyword search. In order to speed-up the keyword search,
it computes the so-called ``inverted file'' for the XML
document -- a set of inverted element lists -- and stores the contents
within a relational database.

\cite{mlca} presents two approaches to compute the Meaningful Lowest
Common Ancestor (MLCA), a concept similar to the SLCA considered in our
approach. Its first approach allows computing the MLCA with the
help of standard XQuery operations, whereas its second approach
is a more efficient approach that is based on a stack-based algorithm for structural joins.

Similar to XRANK \cite{xrank} is the stack-based approach presented in
\cite{stacklca}. In contrast to the previous stack-based appraoches, the authors
do not used the DeweyID to identify a node and to calculate the
ancestor-descendant or even parent-child relationships, but they propose to use
a combination of preorder position, postorder position, and depth of the node.

XKSearch \cite{xksearch} is an indexed-based approach to compute the LCA. They
store inverted element lists consisting of DeweyIDs of the nodes. They start
searching for the results at node $n$ of the shortest relevant keyword list, and 
they check for the other keyword lists whether the node $l$ being the next node
to the left of $n$ or the node $r$ being the next node to the right of $n$ has a
smaller distance to $n$. Then, they use $n$ and the nearest node ($l$ or $r$) to
compute the LCA.

\cite{multiway} presents an anchor-based approach to compute the SLCA. From the
set of current nodes of each relevant keyword list, they search the so-called
anchor, i.e., that node that is closest to all current nodes. As soon as an
anchor is identified, they try to exchange each node $n_i$ of each other keyword
list $L_i$ by the next node next($n_i$) of $L_i$, in order to check, whether
next($n_i$) is closer to the anchor than $n_i$ and whether next($n_i$) defines a
new anchor.
Finally, the set of anchor nodes form the set of LCA candidates that do not
have another LCA candidate child is then reduced to the set of SLCA nodes.

JDeweyJoin \cite{jdewey} returns the top-k most relevant
results. They compute the results bottom-up by computing a kind of join on the
list of DeweyIDs of the nodes in the inverted element list. Whenever they find a
prefix that is contained in all relevant element lists, the node with this
prefix as ID is a result candidate. In addition, they use a weight function to
sort the list entries in such a way, that they can stop the computation after k
results, returning the top-k most relevant results.

\cite{setintersection} and \cite{setintersection2} belong to the
intersection-based approaches. They present a more efficient, but more
space-consuming approach. The elements of their inverted element lists do not
only contain the nodes that have the keyword as label, but also contain all
ancestor-nodes of these nodes, and for each node, the inverted element lists
contain the ID of the parent node. Therefore, they can compute the SLCAs by intersecting the inverted
element lists of the keywords and by finally removing each result
candidate, the descendant of which is another result candidate.

Like the contributions of the second category, our paper focuses on efficient
result computation. It follows the idea of the intersection-based approaches.
However, different from all other contributions and similar to a prior
approach \cite{boettcher}, instead of computing an XML-index, we compute a DAG-Index.
This helps to compute several keyword search results in parallel, and thereby
speeds-up the SLCA computation. To the best of our knowledge, DAG-Index is the
first approach that improves keyword search by using XML compression before
computing the search index.

\section{Conclusions}\label{sec:conclusions}

We have presented IDCLuster, an indexing and search technique that shares
common sub-trees in order to index and to search redundant data only once.

As our performance evaluation shows, using the DAG-based index of
IDCluster, the intersection-based keyword search algorithms can be significantly
improved, i.e., gain a speed-up up to a factor of more than 2.

Therefore, we consider the idea to cluster repeated data collections to be
a significant contribution to all applications that have to search in large
data collections with high amounts of copied, backed-up or redundant data.

\section*{Acknowledgments}

We would like to thank Wei Wang for helpful discussions and comments during our
project cooperation.

This work has been supported by DAAD under project ID 56266786.

\bibliographystyle{IEEEtran}

\bibliography{DAGKwSearch_upb}

\end{document}